\documentstyle[12pt]{article}
\newcommand{\PSbox}[3]{\mbox{\includegraphics{#1}\hspace{#2}\rule{0pt}{#3}}}
\begin{document}
\begin{titlepage}

\begin{flushright}
 IAS-HEP-99-39\\
 PUPT-1783
\end{flushright}

\bigskip
\bigskip
\begin{center}
\LARGE
{\bf Minimal Potentials with Very Many Minima} \\
\bigskip

\bigskip

\normalsize
\large
{Marin Solja\v{c}i\'{c} \footnote{Princeton University, Princeton, USA; 
{\sl soljacic@princeton.edu}}\\
Frank Wilczek \footnote{Institute for Advanced Study, Princeton, USA;
{\sl wilczek@ias.edu}}\\}
\normalsize
\bigskip

\bigskip

\bigskip

\large{\bf Abstract}
\end{center}
\bigskip

We demonstrate, by construction, 
that simple renormalizable matrix potentials with $S_N$, as opposed
to $O(N)$, symmetry can exhibit an exponentially large number 
of inequivalent deep local minima.

\thispagestyle{empty}
\end{titlepage}

There are many situations where behavior of great complexity arises,
or is thought to arise, from simple underlying equations.  Extensively
studied cases include chaos, turbulence, and spin glasses.  Chaos
and turbulence involve long-term dynamics and extended spatial
structures, 
while spin glasses involve an element
of randomness.   Here we will analyze a much simpler case (the
simplest known to us) involving
a static, deterministic, and very symmetrical system, wherein simple
equations exhibit quite a complicated space of solutions. In particular,
we present a simple class of potentials in $n$-component order parameters,
whose number of local minima unrelated by symmetry  grows
exponentially in $n$. Our model
is closely related to ones commonly used 
in studying large $N$ limits of quantum field
theory \cite{coleman}, 
differing only in that the assumption of some continuous 
symmetry among
the fields (e.g., $O(N)$) is replaced by a discrete permutation
symmetry (basically $S_N$).  Of course, it is just such permutation
symmetries which arise in studies of quenched disorder by the replica
method, so there is a close connection to that branch of spin glass
theory \cite{parisi,cugliandolo}.   
In some circumstances the flexibility afforded by imposing less symmetry
might allow  better extrapolations than the traditional one, in the
sense that $1/N$ corrections might be made smaller, and more complex
behaviors captured.

To put the later results in perspective, and to highlight the minimal
requirements for complexity in our framework, let us first consider an
example that does not work.
Suppose that we have
$N$ order parameters
$\phi_i$, for $i=1...N$.  The most general renormalizable (i.e.,
no more than quartic) potential symmetric under the $S_N$ permuting
these parameters and under a change in all their signs simultaneously is 
\begin{eqnarray}\label{pot1}
V(\vec{\phi})=\mu\sum_{i} \phi_i^2 + 
\alpha\sum_{i,j}\phi_i\phi_j + \nonumber 
\beta_1\sum_i \phi_i^4 +
\beta_2\sum_{i,j} \phi_i^3\phi_j \\ +
\beta_3\sum_{i,j} \phi_i^2\phi_j^2 +
\beta_4\sum_{i,j,k} \phi_i^2\phi_j\phi_k +
\beta_5\sum_{i,j,k,l} \phi_i\phi_j\phi_k\phi_l ~.
\end{eqnarray}
Varying with respect to $\phi_a$, we find that for an extremum $\phi_a$ must
obey a cubic equation.
The equation can be written in a form that is the same for 
every value of $a$. That is, for a particular fixed minimum,
the equation can be written as a cubic equation in $\phi_a$,
so that the coefficients of the various terms (when evaluated
as constant numbers for the particular minimum in question)
are the same for all values of $a$. This cubic
equation has at most three real roots.  Of these, at most two are local
minima.  Therefore, for any local minimum, the
different components of the order parameter will take at most two
distinct values.   Thus for large values of $N$ many of the
components will be equal.  Let
us suppose there are $n_1$ components with value $r_1$ and $n_2$
components with value $r_2$, where $n_1 + n_2 = N$, and $n_2 \leq n_1$. 
Then (for given $n_1, n_2$) 
the conditions for an extremum will be two polynomial equations
of degree 3 in the variables $r_1$, and $r_2$.  
In general, these will have at
most 9 solutions.  Taking into account that there are at most two solutions
when $n_2 = 0$, for generic values of the parameters 
$\mu$, $\alpha$, and $\beta_1 ... \beta_5$ in the
potential we readily bound the number of distinct minima
depths by $(9N+4)/2$ for
$N$ even, and $(9N-5)/2$ for $N$ odd.  We
expect that with more care this number could be further 
reduced.  Non-generic values presumably correspond either to
fine tuning of the parameters, which is not physically realistic, or to
enhanced symmetry, which renders mathematically distinct 
solutions physically 
equivalent.  In any case, one does not find here a straightforward
possibility for the sort of exponential growth in the number of
physically distinct minima that we will encounter shortly.  

The next logical step after the vector order parameter is
to consider matrices.  For simplicity we impose that our matrices are 
symmetric, so that $M_{ij}$ and $M_{ji}$ are actually
the same variable. 
Each index runs from 1 to $N$, implying that there
are $N(N+1)/2$ independent order parameters in the
matrix {\bf M}.  We 
define the $a^{th}$ ``row-column'' of a matrix to be 
the union of the $a^{th}$ row, and the $a^{th}$ 
column of the matrix; it is the set of all $M_{ia}$'s
and all $M_{ai}$'s for all $i$'s.

We assume the potential is symmetric under 
{\bf M}$\rightarrow${\bf - M}, and under permutation of the values
of the labels; none
of the row-columns is to be singled out in
any way.  For
example, one can take a matrix, and every time one sees
index 3 in the matrix, replace it with index 7,
and vice-versa. 
Thus entries $M_{37}$, and $M_{73}$ stay
the same, while the entries $M_{33}$ and $M_{77}$
get interchanged, and for all other $i$'s, $M_{i3}$ 
swaps with $M_{i7}$, and $M_{3i}$ swaps 
with $M_{7i}$.  We refer to this
symmetry as the ``row-column exchange symmetry,'' or simply
``exchange symmetry'' of the potential.

Given these constraints, the allowed quadratic terms
in a potential are:
\begin{equation}
M_{ii}M_{ii},M_{ii}M_{ij},M_{ii}M_{jj},M_{ij}M_{ij},
M_{ii}M_{jk},M_{ij}M_{ik},~{\rm and}~M_{ij}M_{kl},
\end{equation}
Here, and hereafter, 
summation over all indices, even if they
are not repeated, is assumed unless explicitly stated
otherwise. 

There are many  allowed quartic terms, and we will not write
them all out here.  But for future reference note 
that terms as highly structured as $M_{ij}M_{jk}M_{kl}M_{li}$, and 
$M_{ii}M_{ij}M_{jk}M_{kl}$ are fair game now.

We will now demonstrate, by explicit construction, 
exponential proliferation of inequivalent local minima in this case.
Our strategy will be to use a subset of the allowed terms to construct
a very simple potential with many
isolated local minima.  These will be equivalent under a symmetry of the
simplified
potential, but not under the smaller symmetry of our
full class of allowed potentials.  Then we shall lift the degeneracy
(and physical equivalence) of these minima in a controlled way by
perturbing with additional allowed terms, in such a way that they
remain local minima.

To begin, we form what we call a plastic-soda-bottle-bottom
potential out of the allowed terms.  (In contrast to the classic
wine-bottle potential, the plastic-soda bottle potential in two
variables has four symmetrically arranged dips.)
This takes the form:
\begin{equation}
V({\bf M})= a \left[ \sum_{i,j}(1-M_{ij}^2) + 
\sum_i(1-M_{ii}^2) \right]^2 +
b \left[ \sum_{i,j} (1-M_{ij}^2)^2 + \sum_i (1-M_{ii}^2)^2 \right] , 
\end{equation}
where $a,b > 0$ are arbitrary, and no summation
over the indices is assumed within the explicitly stated sums.
All the local minima lie at:
\begin{equation} \label{coca_cola}
\left[
\begin{tabular}{cccccc} 
$\pm1$ & $\pm1$ & . & . & . & $\pm1$ \\
$\pm1$ & $\pm1$ & & & & $\pm1$ \\
. & & . & & & . \\
. & & & . & & . \\
. & & & & . & . \\
$\pm1$ & $\pm1$ & . & . & . & $\pm1$ \\  
\end{tabular} \right]  ~.
\end{equation}
They are all related by the accidental symmetry of the 
plastic-soda-bottle potential,
which allows both independent changes in the signs of individual
components and 
interchange of any two components (not just row-columns).

Now we can experiment numerically by adding in more of the 
allowed terms.
The positions of the minima, and their depths, will change
as we vary the amounts of the various small terms we are
adding.  We will be careful that the terms we
are adding are small enough so as not to destabilize any minimum, 
nor change the sign of any of the order parameters at the position
of any of the minima. 
Let us add
terms of the form $M_{ij}M_{jk}M_{kl}M_{li}$ and
$M_{ii}M_{ij}M_{jk}M_{kl}$ with small coefficients,  
with $N=2,3,4,5,6$, and
track 
the depth of each minimum numerically.  Then we can count the number of
distinct numerical values for the potential at the perturbed minima.
Of course, local minima with distinct energies must be physically
inequivalent, {\it i.e}. unrelated by an underlying symmetry.  
The results are exhibited in
Figure 1. 
 
\begin{figure}[h]
\PSbox{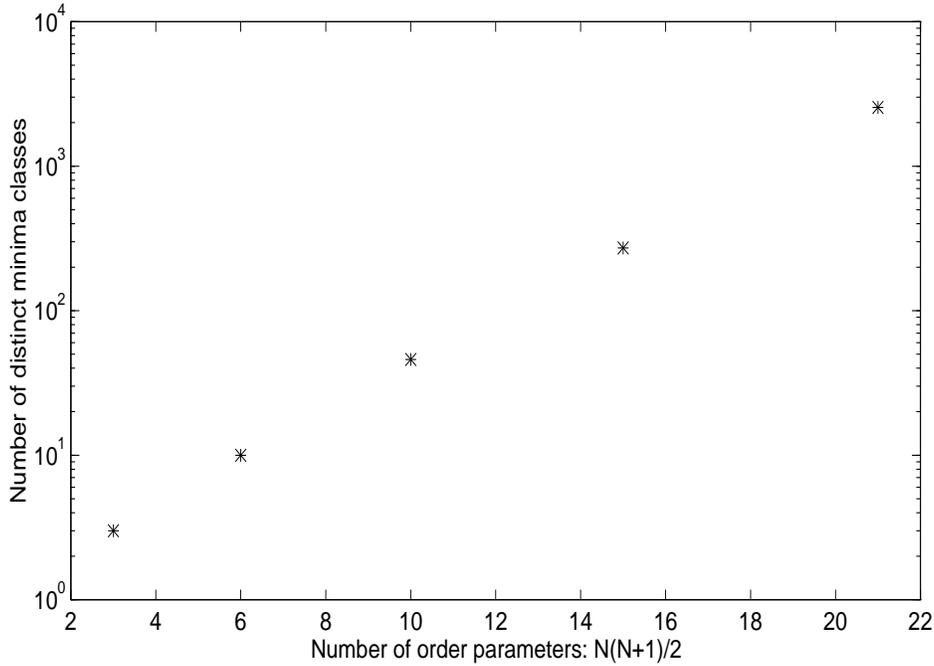 hoffset=12 voffset=-45 
hscale=67.5 vscale=80}{5.4in}{3.5in}
\caption{\footnotesize{On a semi-log plot, the number of
distinct minima classes versus the number of order
parameters appears as a straight line. This is evidence that 
the number of distinct minima
grows exponentially with the number of order parameters.}}
\end{figure}

In Figure 1, it appears that the number of distinct
minima classes grows exponentially
in the number of the order parameters, in response to only these
two particular terms for the perturbations.
Now we shall discuss how this
proliferation can be understood theoretically.

First, let us show
that the number of minima which are not related
by the exchange symmetry, or by the {\bf M}
$\leftrightarrow ${-\bf M} symmetry grows exponentially 
in the number of the order parameters. 
We focus our attention on one very particular
subset of all minima, and prove that the logarithm of the 
number of minima in this subset that are not related 
by any of the allowed symmetries grows quadratically with 
$N$. 
The subset in question consists of all minima
that can be written in the following form:

\begin{equation} \left[
\begin{tabular}{cc}
{\bf B} & {\bf A} \\ 
${\bf A^T}$ & {\bf C} \\ 
\end{tabular} \right] ,
\end{equation}
where when $N$ is even, all matrices {\bf A},
{\bf B}, and {\bf C} have $N/2$ rows and $N/2$ columns; 
and when $N$ is odd, {\bf B} has $(N+1)/2$
rows and $(N+1)/2$ columns, while {\bf C} has $(N-1)/2$
rows and $(N-1)/2$ columns; consequently, {\bf A} has
$(N+1)/2$ rows, and $(N-1)/2$ columns.  
Furthermore, we require that
the matrix {\bf B} has only positive values
on the diagonal (from now on denoted by a $+$,)
while the matrix {\bf C} has only negative values on the
diagonal (from now on denoted by a $-$,) 
while any other entry of {\bf B}, and {\bf C} 
is ``free'' to be either a $+$, or
a $-$.   Note that the logarithm of the 
number of such ``free'' entries grows quadratically in $N$, or
linearly in the number of order parameters, for large $N$. 
Finally, all the entries of the matrix {\bf A} are
fixed; if $N$ is even, all elements on or above the diagonal are
$+$'s, 
while all the elements
below the diagonal are $-$'s; if $N$ is odd,
entry $A_{kl}$ is $+$ if $k \leq l$, and $-$ otherwise.

Concretely, if $N=8$, an element of our subset looks like:

\begin{equation} \left[
\begin{tabular}{cccccccc}
+ & ? & ? & ? & + & + & + & + \\
? & + & ? & ? & - & + & + & + \\
? & ? & + & ? & - & - & + & + \\
? & ? & ? & + & - & - & - & + \\
+ & - & - & - & - & ? & ? & ? \\
+ & + & - & - & ? & - & ? & ? \\
+ & + & + & - & ? & ? & - & ? \\
+ & + & + & + & ? & ? & ? & - \\ 
\end{tabular} \right]  ,
\end{equation}
where $?$ can be either a $+$ or a $-$, as long as it is 
consistent with the requirement $M_{ij}=M_{ji}$;
i.e. the elements below the diagonal are fixed once we
pick the elements above the diagonal. If $N=7$, an element
of our subset looks like:

\begin{equation} \left[
\begin{tabular}{ccccccc}
+ & ? & ? & ? & + & + & +  \\
? & + & ? & ? & - & + & +  \\
? & ? & + & ? & - & - & +  \\
? & ? & ? & + & - & - & -  \\
+ & - & - & - & - & ? & ?  \\
+ & + & - & - & ? & - & ?  \\
+ & + & + & - & ? & ? & -  \\ 
\end{tabular} \right]  ,
\end{equation}
with same requirements as for the case $N=8$.

The reason we focus our attention on this particular
subset is that none of its elements are related 
by the symmetries of our class
of potentials, as we now discuss.  
The proof proceeds in two
steps.  First, ignoring
the existence of the {\bf M}$\leftrightarrow ${\bf - M}
symmetry, we prove that exchange symmetry alone
cannot change one member of subset into another. Then we prove that the 
{\bf M}$\leftrightarrow ${- \bf M} symmetry 
does not cause any further problems.

We propose a ``painting scheme''
to keep track where each entry of the matrix moves during the
exchange process. 
This scheme also makes it easier to visualize what is going on.
Paint each row-column with a different color.
Consequently, each $M_{ij}$ for $i \not= j$ is covered with two layers
of distinct paints; $M_{ii}$ is covered with two layers of the
same paint. Make sure to use ``light colors'' if $+$ is on the 
diagonal entry of the row-column you are painting, and ``dark colors'' if
$-$ is on the diagonal entry. Each particular entry $M_{ij}$ for 
$i \geq j$ is is now labeled uniquely by its two colors;
of course, $M_{ij}$ has the same colors as $M_{ji}$, which
suits us because they are the same variable anyway. 

Say the $2^{nd}$
row-column is yellow, and the $5^{th}$ row-column is green. Exchanging 
indices 2 and 5 makes the $5^{th}$ row-column yellow, and the
$2^{nd}$ row-column green. Using the coloring scheme, it is 
easy to keep track where each particular entry moved during the
exchange. Say the $11^{th}$ row-column was blue initially, 
and we want to know
where the entry $M_{2,11}$ ended up after the exchange; we 
look for the square of the matrix that is covered
precisely by the yellow, and the blue paint, and conclude that
the entry in question is now at the position $M_{5,11}$. 

Note that every entry of the {\bf B} matrix initially
contains only light colors, while the matrix {\bf C} contains
only dark colors.  In contrast, every entry of matrix {\bf A}
is painted with precisely one light, and one dark color.

Now, we start with a matrix ${\bf M_1}$ and permute it into a matrix
${\bf M_2}$ so that both of these matrices are elements of our preferred
subset. First, note that all rows of ${\bf A_2}$ and
${\bf B_2}$ are painted with light colors, while all columns
of ${\bf A_2}$ and ${\bf C_2}$ are painted with darker colors;
this is so because ${\bf B_2}$ has only $+$'s on the diagonal,
while ${\bf C_2}$ has only $-$'s on the diagonal.
Therefore, the set of all entries of ${\bf A_1}$ is exactly
the same as the set of all entries of ${\bf A_2}$; only these
entries are such as to have exactly one light, and one dark color.
Suppose that the light colors we have are: yellow,
orange, red and pink, and suppose $N=8$. Furthermore, suppose 
that ${\bf A_1}$ has
the $1^{st}$ row yellow, the $2^{nd}$ row orange, etc. Since two
entries that were in the same row-column before the exchanges
stay in the same row-column after the exchanges, the only way to get
exactly $4+$'s in the $1^{st}$ row of ${\bf A_2}$ is to have the $1^{st}$
row of ${\bf A_2}$ yellow. This implies that the $1^{st}$ row-column
of ${\bf M_2}$ is yellow. Furthermore, the only way to have 
exactly $3+$'s
in the $2^{nd}$ row of ${\bf A_2}$ is to have the $2^{nd}$ row of
${\bf A_2}$ orange, implying that the $2^{nd}$ row-column of
${\bf M_2}$ is orange, etc. This way we determine the position of all light
colors, and thereby determine uniquely everything about
the matrix ${\bf B_2}$. In a similar manner, we determine 
everything about the matrix ${\bf C_2}$. 
Therefore ${\bf A_1=A_2}$,
${\bf B_1=B_2}$, and ${\bf C_1=C_2}$, so
${\bf M_1=M_2}$, as we sought to prove, since this implies that
there is no symmetry that relates any two elements of this 
particular subset.

The particular
case $N=8$ is just illustrative; everything we said generalizes 
immediately to any even $N$. Furthermore, everything we said
applies with only minor modifications to the case where
$N$ is odd. 

\smallskip     

Now we prove that during the whole process 
of transforming matrix ${\bf M_1}$ into matrix ${\bf M_2}$, one always 
has to multiply the matrix with $-1$ a total of an even number of times. 
The way to see this differs a bit in the case when $N$ is even,
and when $N$ is odd. When $N$ is odd, we have 
to end up with less $-$'s than $+$'s on the diagonal of ${\bf M_2}$,
which is the same as for the diagonal we started with; however,
none of the entries of the diagonal ever moves off the diagonal
during the process.
Similarly, in the case $N$ is even, we have to end up with less $-$'s 
than $+$'s in the matrix ${\bf A_2}$, and we already proved during
Step 1 of this proof that ${\bf A_1}$ consists of the same set of 
elements as ${\bf A_2}$. Therefore, the matrix has to be
multiplied with $-1$ an even number of times during the
process, both when $N$ is odd, and when $N$ is even. 
Since the operation 
of multiplication with $-1$ treats all the elements of the 
matrix indiscriminately, 
it does not matter at all when during the process we perform these
operations; in particular, we could instead perform all of them 
before doing anything else; but then, we 
might as well not do them at all, since multiplying 
the matrix with $-1$ an even number of times leaves 
the matrix unchanged. 

This concludes our proof that the number of local minima of the
special potential that
are unrelated by any symmetry of the general potential grows exponentially
in the number of the order parameters for large $N$.

Physical intuition suggests that unless
two minima have a very good reason to have the same
depths (e.g. an underlying symmetry of the full potential), 
generically one would not expect them to have equal depths. 
Since the potentials of our class support 
an exponentially large number of minima unrelated by symmetry,
we expect that such potentials will generally
have a number of distinct depths at local minima that is exponential
in the number of order parameters, unless the equations that determine
them are insensitive to the symmetry-breaking structure.  That of
course is
the behavior indicated by our numerical work, and it differs markedly
from the earlier, vector case. 
The following consideration
makes it plausible, though it does not prove,  
that the degeneracy among the physically distinct minima, 
which occurs for
our initial plastic-soda-bottle potential, is lifted by
perturbation with certain of the allowed potential terms.    
The point is that the derivative
with respect to $M_{ij}$ of a term like
$M_{ab}M_{bc}M_{cd}M_{da}$, that is 
$M_{ib}M_{bc}M_{cj}$, probes the whole structure of {\bf M\/} in a way
that is significantly different for each value of $ij$.  
Thus, unlike in the vector
case, here the response to the perturbation in principle 
knows enough about (contains
enough independent measures of) the order
parameter to encode its detailed structure.  In the vector case, one
would need to go to {\bf N}$^{th}$ order terms, of the type $\phi_1 \phi_2
... \phi_N$, or higher to encounter similar sensitivity.  

To illustrate this point further, 
we now examine the properties of
some particular cases of our potentials, thus
showing concretely how the various minima become physically
inequivalent.

To keep things as simple as possible, we just add a tiny perturbation
to the initial plastic-soda-bottle potential. 
Because the perturbations are tiny, we 
are justified in evaluating the changes in the potential only to 
the first order; we say that the depth of
each minimum moves by whatever the perturbation we are
adding evaluates to at the original position of the
minimum in question; these positions are given in (\ref{coca_cola}).
To the first order, the degeneracy
can not be broken into an exponentially large number of
minima classes; for example, a quartic term that involves as many as 8 
different indices
can assume at most ${\cal O} (N^8)$ different values when
evaluated at the positions given in (\ref{coca_cola}), since
it is a sum of $N^8$ terms each of which can be either a $+1$, or
a $-1$. Even
if we add all the allowed terms, each multiplied by an arbitrary
tiny coefficient, at lowest order 
we still have at best a power law breaking
of the degeneracy. 

Nevertheless, the number of distinct minima
one can in principle get by analyzing only to the first order is 
quite large, especially if we include many
allowed terms to create the perturbation.  Furthermore,
for small perturbations, the expectation values of different operators
will typically not differ significantly if we evaluate the 
changes in the depths 
only to the first order, as opposed to evaluating them exactly. 
Moreover, in practice we sort the minima
into energy bins of finite width in our plots. If our perturbation
breaks the degeneracy to the first order into say ${\cal O}(N^8)$ 
distinct minima classes and $N=6$, we have in principle up
to $\sim 10^6$ distinct minima. Since our plots
typically involve 200 bins, it does not matter
for the plots that we evaluate the depth changes to the first 
order only instead of calculating them exactly. 

\begin{figure}[ht]
\PSbox{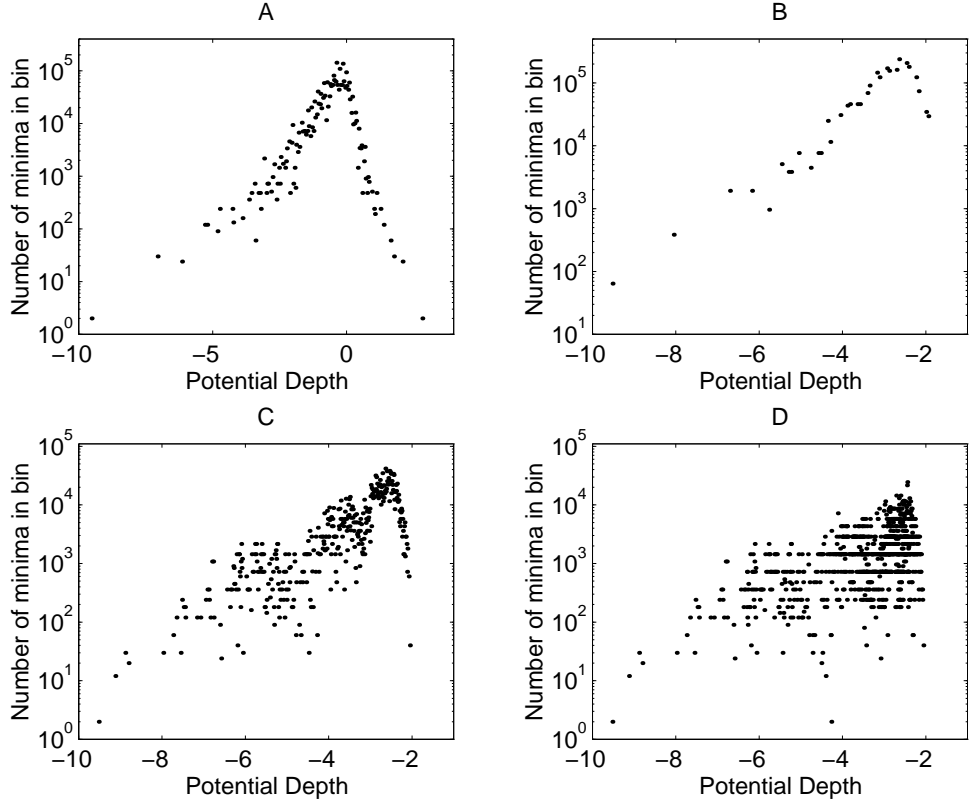 hoffset=-12 voffset=-144 
hscale=75 vscale=75}{5.4in}{4.25in}
\caption{\footnotesize{Plots of bin occupation numbers versus the 
changes in minima depths, evaluated to the first order in the small
perturbations. All plots are for $N=6$. The 
x-axes are in arbitrary units. 
The width of the bins in plots A,B, and 
C is $10^{-2}$ energy units, and in plot D is 
$2.5*10^{-4}$ energy units. The term added in plot A
was $-M_{ii}M_{ij}M_{jk}M_{kl}$, while in plot B it
was $-M_{ij}M_{jk}M_{kl}M_{li}$. To create plot C,
we add 20 different terms, with random coefficients
multiplying them. Plot D is exactly the same as plot
C, except with much higher bin resolution.}}
\end{figure}

Typical results are displayed in Figure 2. 
Plots A and B from that figure demonstrate that one can get quite a rich
structure by using only a few of the allowed terms. 
Furthermore, the breaking of degeneracy is quite
extensive  even when we work to first order only. 
When we include more than one perturbative term, the 
degeneracy breaking is 
even bigger, producing quite a rich structure even at first order. 
This is visible in plots C and D of Figure 2, where we included 20 
of the allowed terms, with random coefficients multiplying them. The plot 
D has a very high resolution of almost 40000 bins for the 
whole plot; both plots are for exactly the same potential.
Note that in these plots we count the total number of minima, 
so that minima are counted as distinct even if they are related
by a symmetry. Thus, much of the degeneracy is intrinsic,
and will not be broken in any order of approximation. 

An example of the general sort of structure described here arises in
the analysis of QCD with many flavors of quarks at high density.  For
three flavors the color-flavor locking condensate takes the form
\cite{arw} 
\begin{equation} \langle q^\alpha_a q^\beta_b \rangle ~=~
U^\alpha_\gamma U^\beta_\delta (\kappa_1 \delta^\gamma_a 
\delta^\delta_b + \kappa_2 \delta^\gamma_b \delta^\delta_a )~,
\end{equation}
where the Greek indices refer to color and the Latin to
flavor.  For present purposes we are suppressing various inessential
complications (spin, chirality, momentum dependence), and emphasizing
the existence of the matrix degree of freedom $U$, which parameterizes
the degenerate vacua associated with the spontaneous symmetry breaking
$SU(3)_{\rm color} \times SU(3)_{\rm flavor} \rightarrow SU(3)_{\rm
color+flavor}$.

It appears that for $3k$ flavors the favored condensation is repeated
color-flavor locking \cite{sw}.  Thus we start with the {\it ansatz\/}
\begin{equation}
\label{multiCFL}
\langle q^\alpha_a q^\beta_b \rangle ~=~ \sum_{i=1}^k
U^{(i)\alpha}_\gamma U^{(i)\beta}_\delta (\kappa_1
\delta^\gamma_{a-3i+3} \delta^\delta_{b-3i+3} + \kappa_2
\delta^\gamma_{b-3i+3}\delta^\delta_{a-3i+3} )~, \end{equation}
 corresponding to the symmetry breaking $SU(3)_{\rm color} \times
SU(3k)_{\rm flavor} \rightarrow SU(3)_{\rm color+diagonal} \times
S_k$.  The residual $SU(3)$ acts on the flavor indices in blocks of 3,
while the permutation symmetry $S_k$ implements interchanges of the
blocks.

Now the question arises how the energy depends on the relative
alignment of the $U^{(i)}$.  Non-trivial relative alignments violate
the permutation symmetry.  We will not attempt here to determine
whether this actually occurs in the ground state, or in other
low-lying states, but we do want to point out that to analyze this
question one would need to consider potentials resembling those
discussed above, featuring permutation rather than rotation symmetry
in internal space.  This case is intermediate in complexity between the
vector and matrix cases discussed above, in that the permutation acts
on a single index (as in the vector case), but the objects being
permuted are chosen from a complicated manifold, rather than being a
simple choice of sign.  Symmetry breaking correlations of the type
$\langle U^{(i)} U^{(j)} \rangle \sim M^{(ij)}$ could produce an
effective matrix structure in the permutation index.

We acknowledge useful discussions with Shiraz Minwalla of 
Princeton University.

\end{document}